\documentclass[%
 reprint,
 amsmath,amssymb,
 aps,
]{revtex4-2}

\usepackage{graphicx}
\usepackage{dcolumn}
\usepackage{bm}

\usepackage{braket}
\usepackage[colorlinks,linkcolor=blue,anchorcolor=blue,citecolor=blue,urlcolor=blue]{hyperref}

\begin{document}

\preprint{APS/123-QED}

\title{Quantum Hashing via Constrained Rydberg Many-Body Dynamics}

\author{Han-Chao Chen$^{1,2}$}
\thanks{These authors contributed equally to this work.}
\author{Xin Liu$^{1,2}$}
\thanks{These authors contributed equally to this work.}
\author{Zheng-Yuan Zhang$^{1,2}$}
\author{Dong-Sheng Ding$^{1,2}$}
\email{dds@ustc.edu.cn}
\author{Bao-Sen Shi$^{1,2}$}

\affiliation{$^1$Laboratory of Quantum Information, University of Science and Technology of China, Hefei 230026, China.}
\affiliation{$^2$Anhui Province Key Laboratory of Quantum Network, University of Science and Technology of China, Hefei 230026, China.}

\date{\today}

\begin{abstract}
In this Letter, we show that constrained many-body dynamics in Rydberg atom arrays naturally gives rise to a quantum hashing mechanism. By encoding ternary strings into deterministic trajectories in the state space, the classical information space is mapped onto a quantum state ensemble in the Hilbert space with an induced geometric structure. Statistical analysis reveals that this ensemble exhibits high probability near-orthogonality, random-like distribution, and broad geometric coverage. These geometric features naturally give rise to the essential cryptographic properties of quantum hashing, including low collision probability, one-wayness, tamper sensitivity, and privacy preservation. Our results demonstrate that the cryptographic functionality of quantum hashing need not rely on deliberately engineered algorithms, but can instead emerge naturally from constrained many-body dynamics, identifying quantum dynamics itself as a physical resource for cryptographic information processing.
\end{abstract}

\maketitle


\textit{Introduction.---} Information spreading in quantum many-body systems is a central topic in statistical physics and quantum information science \cite{1,2,3,4}. Under many-body interactions, local information can rapidly propagate across an exponentially large Hilbert space, such that the initial information, while preserved in principle, becomes difficult to recover through local observations or simple measurements \cite{2,5,6,7,8,9,10,11,12,13,14}. This process underlies a variety of phenomena, including thermalization, quantum chaos, and the generation of pseudorandom quantum states \cite{15,16,17,18,19,20}. Considerable effort has been devoted to understanding how local quantum dynamics gives rise to complex information structures \cite{21,22,23,24}. However, existing studies have primarily focused on how spreading dynamics conceal information and have largely regarded the resulting complexity as an obstacle to controllable information processing, rather than exploring it as a potentially functional information processing resource \cite{25,26,27,28}.

Cryptography provides an intriguing context in which to revisit this issue \cite{29,30,31}. Classical hashing and its quantum extensions constitute fundamental cryptographic primitives \cite{32,33,34,35,36,37,38,39}. An effective hash function is expected to exhibit several key properties, including low collision probability, the avalanche effect, and one-wayness \cite{40,41}. In recent years, various quantum hashing schemes have been proposed and applied to quantum authentication, quantum communication, and information security \cite{42,43,44,45,46}. However, these approaches typically rely on carefully designed quantum circuits and specifically engineered encoding protocols, with their cryptographic properties arising from deliberate construction \cite{34,36,37,38,47,48}. Given the striking similarities between the requirements of quantum hashing and the characteristics of information spreading, the possibility that cryptographic functionality may emerge directly from quantum many-body dynamics without deliberately designed encryption protocols remains largely unexplored.

In this work, we encode classical ternary strings into deterministic dynamical trajectories based on the many-body dynamics arising from the alternating interplay of blockade, antiblockade, and facilitation mechanisms in Rydberg atom arrays, thereby establishing a dynamical mapping from classical information to many-body quantum states and revealing the geometric structure of the mapped state space. We show that quantum hashing emerges naturally from this geometry induced by the underlying dynamics, and investigate its collision probability, information recoverability, tamper sensitivity, and privacy preserving capability. Our results demonstrate that cryptographic functionality can emerge directly from quantum many-body dynamics, establishing a connection between quantum dynamics and cryptographic primitives \cite{49,50,51,52}.

\begin{figure*}[ht]
    \centering
    \includegraphics[width=1.0\linewidth]{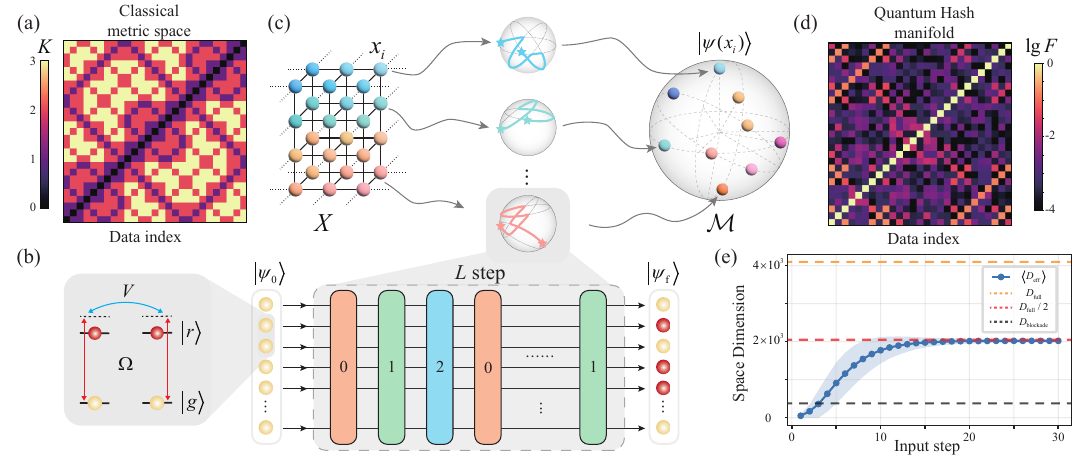}
    \caption{\textbf{Quantum hashing protocol and information space structure.} (a) Classical metric space. A pairwise Hamming distance matrix of ternary data arranged in Gray code. (b) One-dimensional Rydberg atom arrays and schematic diagram of the quantum hashing process. Ternary symbols $\{0,1,2\}$ are mapped to the quantum hashing elementary evolution operator. Under the action of a sequence of length $L$, the initial state $\ket{\psi_0}$ evolves into the final quantum hashing state $\ket{\psi_f}$. (c)  Reconstruction of ternary information space to Hilbert space. Each ternary string $x_i$ in the classical space $X$ uniquely determines a deterministic dynamical trajectory, which is mapped to a single quantum hashing state $|\psi(x_i)\rangle$ in the projected Hilbert space $\mathcal M$. (d)  Quantum Hashing manifold. The pairwise logarithmic fidelity $\lg F$ matrix after mapping. (e) Average effective dimension $\langle D_{\text{eff}} \rangle$ as a function of the input step number. The blue shaded region denotes the standard deviation. Dashed lines represent the full Hilbert space dimension $D_{\text{full}}$, half of the full Hilbert space dimension $D_{\text{full}}/2$, and the Rydberg blockade subspace dimension $D_{\text{blockade}}$, respectively.}
    \label{fig1}
\end{figure*}

\begin{figure}[htpb]
    \centering
    \includegraphics[width=1.0\linewidth]{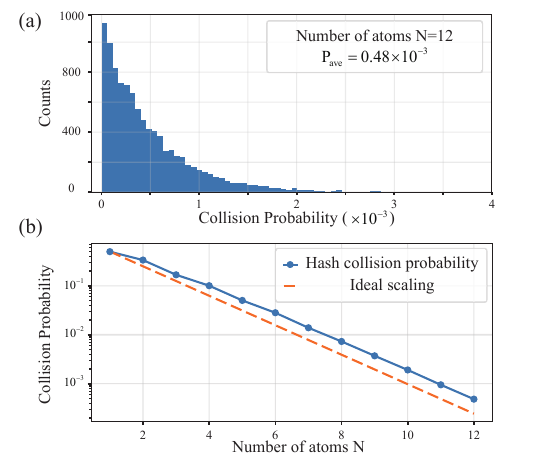}
    \caption{\textbf{Hashing collision probability.} (a) Distribution of hashing collision probabilities obtained from Monte Carlo simulations with $10^4$ samples for a system of $N=12$ atoms. The corresponding average collision probability is calculated to be $P_{\text{ave}}=0.048\%$. (b) Average hashing collision probability as a function of the number of atoms $N$, with the vertical axis shown on a logarithmic scale. The blue solid line represents the collision probability obtained from numerical statistics, while the orange dashed line denotes the scaling behavior expected for an ideal hash function, $P_{\text{coll}}=1/2^N$.}
    \label{fig2}
\end{figure}

\textit{Model and Quantum Hashing.---} We begin by considering the classical information space $X=\{0,1,2\}^L$, consisting of ternary strings of length $L$, where each element $x=(x_L,x_{L-1},\cdots,x_1)$ uniquely represents a classical input. The geometry of this space is characterized by the Hamming distance \cite{66,67}:
\begin{equation}
K(x,y)=\sum_{i=1}^{L}\delta_{x_i\neq y_i},
\end{equation}
which counts the number of symbols at which two strings differ. The pairwise Hamming distance matrix for the classical information space with $L=3$ are show in Fig.~\ref{fig1}(a). To visualize its local geometric organization, the ternary strings are reordered according to a Gray code sequence, such that neighboring indices correspond, whenever possible, to strings differing by a single symbol. The resulting matrix exhibits a pronounced block structure, reflecting the locality of the classical information geometry. This discrete metric space possesses an additive geometric structure, in which each symbol contributes independently to the total distance without introducing any nonlocal correlations beyond sitewise differences. Accordingly, the classical input space is described by the metric measure space $(X,K,\mu_X)$, where $\mu_X(x)=3^{-L}$ is the uniform probability measure on $X$, reflecting the equal likelihood of all classical inputs \cite{68,69,70}.

To encode classical information into quantum states through many-body dynamics, we introduce a set of dynamical generators $\mathfrak A=\{O_F,O_B,O_P\}$, corresponding respectively to blockade excitation, pair antiblockade, and facilitation processes. Defining the Rydberg occupation operator $n_i=\ket{r_i}\bra{r_i}$, the ground state projector $P_i=1-n_i$, the local spin flip operator $\sigma_i^x=\ket{g_i}\bra{r_i}+\ket{r_i}\bra{g_i}$, and the raising and lowering operators $\sigma_i^+=\ket{r_i}\bra{g_i},\quad \sigma_i^-=\ket{g_i}\bra{r_i}$, the three generators are given by:
\begin{gather}
    O_B=\sum_iP_{i-1}\sigma_i^xP_{i+1},\\
O_P=\sum_iP_{i-1}
\left(
\sigma_i^+\sigma_{i+1}^+
+\sigma_i^-\sigma_{i+1}^-
\right)
P_{i+2},\\
O_F=\sum_i\Pi_i\sigma_i^x,
\end{gather}
where $\Pi_i=n_{i-1}P_{i+1}+P_{i-1}n_{i+1}$ projects onto configurations containing exactly one excited nearest neighbor. The system evolves under the effective Hamiltonian \cite{53,54,55}:
\begin{equation}
H(t)=\sum_{\mu\in\{B,P,F\}}\lambda_\mu(t)O_\mu,
\end{equation}
where the control function $\lambda_\mu(t)\in\{0,1\}$ specifies which constrained dynamical process is activated during each time interval. Each elementary evolution operator is then associated with one ternary symbol:
\begin{equation}
\{\mathcal U_F,\mathcal U_B,\mathcal U_P\}
\rightarrow
\{0,1,2\},
\end{equation}
where $\mathcal U_\mu=e^{-i\tau_\mu O_\mu},\quad\mu\in\{F,B,P\}$, with $\tau_\mu$ chosen as one-half of the oscillation period of the corresponding elementary excitation process. For any classical input $x\in X$, the corresponding dynamical trajectory is $\mathcal U(x)=\prod_{j=0}^{L-1}\mathcal U_{x_{L-j}}$, which prepares the quantum state
$\ket{\psi_{\text{f}}(x)}=\mathcal U(x)\ket{\psi_0}$ as show in Fig.~\ref{fig1}(b). The system is initialized in the product state $\ket{\psi_0}=\ket{g}^{\otimes N}$. Since the facilitation process requires a pre-existing Rydberg excitation, it follows that $\mathcal U_F\ket{\psi_0}=\ket{\psi_0}$. Consequently, the evolution operator associated with the symbol $\{0\}$ leaves the initial state unchanged, naturally ensuring a unique canonical representation of every encoded sequence.

The above protocol thus realizes a quantum hashing map emerging directly from constrained many-body dynamics:
\begin{equation}
h:\mathbb Z_3^L
\rightarrow
P(\mathcal H),
\end{equation}
where $P(\mathcal H)$ denotes the projective Hilbert space, whose elements are rays representing pure quantum states modulo an overall global phase. As illustrated in Fig.~\ref{fig1}(c), each classical ternary string uniquely specifies a deterministic dynamical trajectory and is consequently mapped to a single quantum hashing state in the projective Hilbert space. The resulting quantum hashing space is defined as:
\begin{equation}
\mathcal M=h(X)
=
\left\{
|\psi(x)\rangle\,\middle|\,x\in X
\right\}
\subset
P(\mathcal H).
\end{equation}
Since the classical input space is equipped with the uniform probability measure, the hashing map naturally induces the pushforward measure $\mu_M=h_*\mu_X$ \cite{71}, which characterizes the statistical distribution of classical inputs over the quantum hashing manifold. The space is therefore described by the metric measure space $(\mathcal M,d_{\mathrm{FS}},\mu_{\mathcal M})$, where the geometry is determined by the Fubini-Study distance \cite{71,72}:
\begin{equation}
d_{\mathrm{FS}}(\psi_i,\psi_j)
=
\arccos
\braket{\psi_i|\psi_j},
\end{equation}
which provides the natural geometric measure of statistical distinguishability between pure quantum states. Equivalently, the local similarity between two hashing states can be quantified by their fidelity, $F=|\braket{\psi_i|\psi_j}|^2$.

The dynamical mapping fundamentally reconstructs the geometry of the information space. Rather than preserving the local neighborhood structure defined by the Hamming metric, it transforms the discrete classical space into a quantum statistical manifold governed by many-body interference and quantum correlations. As illustrated in Fig.~\ref{fig1}(d), the pairwise logarithmic fidelity $\lg F$ matrix, displayed using the same ordering as in Fig.~\ref{fig1}(a), exhibits a strikingly different pattern from its classical counterpart, the pronounced block structure disappears and nearly all off-diagonal elements become strongly suppressed. Consequently, classical strings that are close in Hamming space are mapped to quantum states that are widely separated in projective Hilbert space and, with high probability, become nearly maximally distinguishable. This geometric reconstruction underlies the strong separability of the quantum hashing states and provides the geometric origin of the exponentially suppressed collision probability.

\begin{figure*}[htpb]
    \centering
    \includegraphics[width=1.0\linewidth]{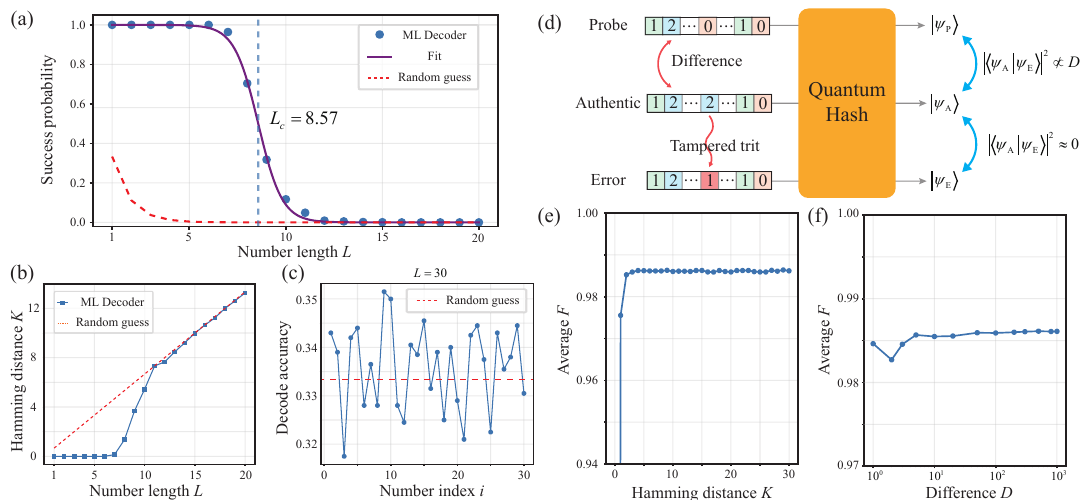}
    \caption{\textbf{Cryptographic properties of quantum hashing.} (a) Random Forest machine learning (ML) decoder success probability versus input length $L$. The blue scatter points represent the numerical data, while the purple solid line denotes the logistic function fitting curve $P_{\rm fit}=\frac{1}{1+\exp[(L-L_c)/k]}$. The red dashed line indicates the baseline corresponding to random guessing by the decoder, and the blue dashed line marks the fitted critical length $L_c$. (b) Hamming distance between the decoded output string and the true input string as a function of the input length $L$. The red dashed line indicating the random guessing limit, $K=2L/3$. (c) Distribution of decoding accuracy for each trit position in the input string when the input length is $L=30$. The horizontal dashed line represents the baseline accuracy of random guessing, $1/3$. (d) Schematic illustration of tamper sensitivity and privacy protection. A single trit modification produces a nearly orthogonal hashing state, while probe state overlap shows no clear dependence on input distance. (e) The average fidelity between quantum hashing states generated from input sequences separated by different Hamming distances. (f) The numerical difference between the authentic input and the probe input, as a function of the average fidelity. }
    \label{fig3}
\end{figure*}

\textit{Spreading Dynamics and Hashing Collision.---} Collision probability is one of the most important metrics for evaluating a hash function. The information geometric picture developed above suggests that the collision probability is fundamentally determined by the geometric distinguishability between quantum hashing states in the Hilbert space. Therefore, achieving low collision probability requires that states generated from different classical inputs are sufficiently dispersed over the accessible Hilbert space, thereby enhancing their mutual distinguishability. To quantify this behavior, we perform Monte Carlo simulations on a system of $N=12$ Rydberg atoms using $10^4$ randomly generated valid input sequences\cite{56,57}.

We first investigate the spreading dynamics of quantum states in Hilbert space during the hashing process. The effective dimension is defined as $D_{\rm eff}=1/\sum_n p_n^2$, where $p_n$ denotes the probability distribution of the quantum state in the computational product state basis \cite{58,59}. The average effective dimension as a function of the input length $L$ is shown in Fig.~\ref{fig1}(e). The sufficiently large exploration of Hilbert space provides a structural basis for enhanced distinguishability between quantum states, making outputs corresponding to different inputs more separable in a statistical sense. Based on this observation, we analyze the fidelity distribution between quantum hashing states generated from random inputs, as shown in Fig.~\ref{fig2}(a). The overwhelming majority of pairwise overlaps are concentrated near zero, yielding an average collision probability of only $P_{\rm ave}=0.048\%$. This demonstrates that distinct classical inputs are typically mapped to nearly orthogonal quantum states, thereby ensuring strong collision resistance.

To further investigate the scaling behavior of collision probability with system size, we compute the average collision probability as a function of the number of atoms, as shown in Fig.~\ref{fig2}(b), with the vertical axis plotted on a logarithmic scale. The numerical results closely follow the theoretical prediction for an ideal quantum hash function \cite{32,34,73}, $P_{\rm coll}=1/D$, where $D=2^N$ denotes the Hilbert space dimension. This agreement demonstrates that the collision probability decreases exponentially with system size, confirming the strong collision resistance of the protocol. The observed scaling reflects the rapid growth of the accessible Hilbert space with system size, which continuously enhances the distinguishability between quantum hashing outputs.

\textit{Cryptographic Properties of the Quantum Hashing.---} In addition to a low collision probability, a desirable hash function should possess several fundamental cryptographic properties, including one-wayness, sensitivity to input tampering, and privacy preservation. We first examine the one-wayness of the quantum hashing proposed here. For a given quantum hashing state, we use its probability distribution in the product state basis as the input feature and construct machinelearning (ML) decoders based on Random Forest (RF), Multi-Layer Perceptron (MLP), and k-nearest-neighbor (KNN) algorithms to recover the original ternary sequence \cite{60,61,62,63,64}. All three decoding methods converge to qualitatively identical results, indicating that the observed behavior is independent of the specific choice of decoder. The Random Forest results are shown in Fig.~\ref{fig3}, which we use for the following analysis.

The probability of exact recovery as a function of the input length $L$ is shown in Fig~\ref{fig3}(a), where exact recovery means that every trit is correctly predicted. As $L$ increases, the success probability exhibits a clear crossover behavior and is well fitted by the logistic function $P_{\rm fit}=\frac{1}{1+\exp[(L-L_c)/k]}$ \cite{65}, with parameters $L_c=8.570$ and $k=0.618$. This crossover is closely related to the size of the training dataset. When $L<L_c$, the training set nearly covers the entire input space, allowing the decoder to achieve a high recovery accuracy. In contrast, when $L>L_c$, the size of the input space, $3^L$, exceeds the training set size, causing the success probability to rapidly approach the random-guessing limit. Consistently, changing the size of the training dataset leads to a corresponding shift of the crossover point $L_c$. More detailed information is provided in Fig. \ref{fig3}(b), where the Hamming distance $K$ of the decoded sequence, gradually approaches the random-guessing limit $2L/3$. Figure \ref{fig3}(c) further shows the prediction accuracy for each digit position at $L=30$. All positions fluctuate around $1/3$, fully consistent with random guessing. These results demonstrate that even when the complete observable probability distribution of the hashing state is provided as input, the original sequence remains difficult to reconstruct, indicating a strong one-way property.

We next investigate two additional cryptographic properties of the proposed quantum hashing protocol, namely tamper sensitivity and privacy preservation, illustrated schematically in Fig.~\ref{fig3}(d). To quantify the tamper sensitivity, we calculate the average fidelity between quantum hashing states generated from input sequences separated by different Hamming distances, as shown in Fig.~\ref{fig3}(e). Even a single trit modification causes the output fidelity to drop sharply, and the fidelity continues to decrease as the Hamming distance increases before rapidly approaching a low overlap plateau. This pronounced avalanche behavior demonstrates that local perturbations of the classical input are efficiently amplified by the constrained many-body dynamics, leading to globally distinguishable quantum hashing states and enabling reliable detection of input tampering \cite{74,75}.

Privacy preservation is examined by analyzing whether the similarity between classical inputs can be inferred from their corresponding quantum hashing states. The average fidelity as a function of the numerical difference between the authentic input and a probe input is presented in Fig.~\ref{fig3}(f). Despite the increasing difference between the two inputs, the output fidelity exhibits no clear monotonic dependence and instead quickly fluctuates around a uniformly low value. This indicates that proximity relations in the classical information space are not preserved under the quantum hashing mapping, and the similarity between classical inputs is therefore concealed in the quantum output space. Consequently, an adversary cannot infer how close a trial input is to the authentic one from the corresponding quantum hashing states, demonstrating a desirable privacy preserving capability.

\textit{Conclusion and outlook.---} This work reveals that quantum hashing can emerge naturally as an information functionality of constrained quantum many-body dynamics, and establish a dynamical framework for realizing quantum hashing based on Rydberg atom arrays. From an information geometric perspective, we develop a unified description of the hashing process and show that constrained many-body dynamics not only maps classical information to quantum states, but also fundamentally reshapes the geometry of the information space, transforming the additive geometry of a discrete classical space into a quantum information geometry governed by many-body interference and quantum correlations. Through this geometric reshaping, distinct classical inputs naturally evolve into nearly orthogonal quantum states, enabling a low collision compression from a $3^{30}$-dimensional classical input space to a $2^{12}$-dimensional Hilbert space without relying on engineered random circuits or dedicated cryptographic protocols. As a consequence, desirable cryptographic properties including low collision probability, strong avalanche behavior, tamper sensitivity, and privacy preserving capability emerge intrinsically from the underlying dynamics. Our results demonstrate that constrained quantum many-body dynamics is not only a powerful platform for quantum simulation, but also a versatile physical resource for information processing. By establishing a direct connection between quantum many-body dynamics and quantum hashing, this work provides a new perspective on the interplay between quantum dynamics, information geometry, and cryptographic functionality, and opens promising opportunities for exploring emergent information functionalities in larger quantum simulators, open quantum systems, and more general classes of constrained dynamics.

We acknowledge funding from the National Key R and D Program of China (Grant No. 2022YFA1404002), the National Natural Science Foundation of China (Grant Nos. T2495253, 61525504, 61435011).

\nocite{*}

\bibliography{citation}

\end{document}